\documentclass[reprint,superscriptaddress,groupedaddress,amsmath,amssymb,aps]{revtex4-2}
\usepackage{multirow}
\usepackage{graphicx}
\usepackage{dcolumn}
\usepackage{bm}
\usepackage{xcolor}
\usepackage{amsmath}

\begin{document}

\preprint{APS/123-QED}

\title{Vibrational Nonlinear Response of Complex Molecules:\\ Nature and Measurements in the THz Range}

\author{Aleksandra Nabilkova}
\author{Mikhail Guselnikov}%
\author{Azat Ismagilov}
\author{Maksim Melnik}
\author{Sergey Kozlov}
\author{Anton Tcypkin}
 \email{tsypkinan@itmo.ru}
\affiliation{%
 Research and Education Center of Photonics and Optical IT, ITMO University, St. Petersburg, Russia 
}%
\affiliation{
 Laboratory of Quantum Process and Measurements, ITMO University, St. Petersburg, Russia.
}%

\date{\today}

\begin{abstract}
The nonlinear optical response of materials under high-intensity electromagnetic fields is key to advancing THz optical technologies. This study introduces a theoretical approach to estimate the nonlinear refractive index coefficient $n_2$ of complex molecules by summing contributions from individual vibrational bonds, which are extracted from structurally simpler molecules predominantly consist of corresponding type of bond. Using water, $\alpha$-pinene, and CO$_2$ as reference materials, we predict $n_2$ for isopropanol as 4.7$\times$10$^{-9}$ cm$^2$/W. Z-scan measurements with pulsed THz radiation yield $n_2$ = 2.5$\pm$0.5$\times$10$^{-9}$ cm$^2$/W, validating our model. This bond-based decomposition links microscopic bond vibrations and macroscopic nonlinear coefficients, offering a framework for understanding molecular nonlinear optics and guiding the design of THz nonlinear materials.
\end{abstract}

\maketitle


\section{\label{sec:level1}Introduction}
The intensive study of nonlinearity in the terahertz (THz) range was driven by both recent spread of high-power THz pulse sources and theoretical prediction of the giant nonlinear refractive index coefficient for some materials that several orders of magnitude larger than in infrared (IR) and visible (VIS) ranges \cite{dolgaleva2015prediction}. Later, several independent groups experimentally confirmed that prediction for crystals, vapors and liquids \cite{tcypkin2021giant,tcypkin2019high, novelli2020nonlinear, francis2020terahertz, guo2020research, rasekh2021terahertz}. Estimates and experimental findings reveal that liquids in the THz range exhibit nonlinear refractive indices approximately 2–3 orders of magnitude higher than those of crystals \cite{tcypkin2021giant,tcypkin2019high, novelli2020nonlinear, francis2020terahertz}, thereby amplifying interest in the detailed investigation of their nonlinear properties and potential applications.

The interaction of THz radiation with liquids enables a variety of applications across scientific and industrial fields. The data on the linear optical characteristics of many liquids, especially alcohols \cite{sani2016spectral}, in the THz frequency range are of importance for medicine, non-destructive testing \cite{liu2013real}, chemistry \cite{funkner2012watching}, environmental monitoring (assessment of water quality and pollution levels) \cite{litra2024propanol}. Polar liquids strongly interact with THz waves via intermolecular hydrogen bonds \cite{laib2010terahertz}, while non-polar liquids exhibit weaker interactions \cite{jepsen2008characterization}. This contrast allows classification, such as detecting alcohols in fuel solutions \cite{lapuerta2020determination}. Understanding of liquid behavior at THz frequencies helps in the design of advanced materials, including those used in coatings, emulsions, and nanofluids \cite{pogosian2024fabrication,frenzel2023nonlinear}.

Since the resonant bands of liquid molecules in the THz range are predominantly of vibrational origin, the theoretical and experimental investigation of the optical properties and temporal dynamics of vibrational bonds is a critical fundamental research area \cite{zhao2021molecular,guselnikov2023}. A comprehensive understanding of these processes could enable precise control over nonlinear optical phenomena occurred in THz range: as an example, controlling the THz spectra of plasma radiation in liquids and THz pulses energy transfer during ultrafast optical processes \cite{chen2023plasma}. Another important direction having required knowledge about THz nonlinear properties of liquids is studying of the collective polaron behavior of electrons and the oriented molecular cloud. This phenomenon arises due to the reorientation of dipolar solvent molecules in response to the generation of free electrons during the ionization of polar liquids such as isopropanol \cite{runge2023nonlinear}. Modeling these nonlinear polaronic interactions necessitates understanding of solvent nonlinearity. According to the first theoretical estimations, the nonlinear refractive index coefficient of alcohols is 1-2 orders of magnitude greater than that of water \cite{tcypkin2021giant}. However, nonlinearity of alcohols is a poorly studied area of knowledge \cite{wang2005vibrational} and such large values have not been confirmed experimentally.

Moreover, giant nonlinear refractive index coefficient of alcohols enables the observation of nonlinear optical effects, including self-focusing, frequency shifts, the Kerr effect, and higher harmonic generation in the THz frequency range, at laser intensities 2–4 orders of magnitude lower than those typically required in the visible and infrared spectral regions \cite{zhao2021molecular,zhang2022terahertz,articleWang}. In turn, these phenomena can be used to create such optical scheme units as ultrafast optical switches, THz-based imaging systems, frequency converters limiters, amplifiers, transistors and potential memory devices for a future investigation of optical computer \cite{nabilkova2023controlling,nabilkova2024transmission}. 

This study reveals novel theoretical approach of decomposing the nonlinear response of a complex molecule into contributions from its constituent chemical bonds by leveraging the known nonlinear properties of structurally simpler molecules containing corresponding individual bonds. We apply this methodology to isopropanol, estimating its nonlinear refractive index $n_2$ as the sum of contributions from O-H, C-H, and C-O bonds. The theoretical prediction of $n_2$ = 4.7$\times$10$^{-9}$ cm$^2$/W is experimentally validated using Z-scan measurements with a pulsed THz radiation source, yielding $n_2$ = 2.5$\pm$0.5$\times$10$^{-9}$ cm$^2$/W. The agreement supports the feasibility of evaluating the nonlinear refractive index of various materials (such as alcohols and other organic compounds) without direct experimental measurements. This approach may facilitate the design and optimization of nonlinear materials for THz-based sensing, ultrafast optical switching, and frequency conversion.

\section{Isopropanol's Nonlinear Refractive Index Coefficient of vibrational nature}\label{theor}
The study by Dolgaleva et al. \cite{dolgaleva2015prediction} introduced a theoretical framework for evaluating nonlinear refractive index coefficient $n_{2}$ of crystalline media in the THz range, based on well-characterized physical and optical properties of the materials. This theory attributes the THz nonlinearity in media predominantly to a vibrational response arising from stretching fundamental vibrations, which surpass the contributions of electronic and orientational nonlinearities. The model posits that the main mechanism contributing to nonlinear refractive index change of media in the THz range is low-inertia anharmonical molecular vibrations. Subsequently, this approach was expanded for the case of water \cite{tcypkin2019high,tcypkin2021giant}, and its applicability was experimentally confirmed \cite{nabilkova2023controlling}. The same model has also been employed and further validated by other research groups \cite{hemmatian2023simplified,rasekh2020propagation}. For a medium dominated by a single vibrational resonance with fundamental frequency $\omega_{0}$, which is substantially higher than $\omega$, the vibrational contribution to $n_{2}$ can be described through well-established liquid properties or easily measurable ones.

\begin{eqnarray}
\begin{aligned}
 &\tilde{n}_{2} \equiv \operatorname{Re}\left[\tilde{\bar n}_{2}^{\omega \ll \omega_{0}}\right] \approx \frac{\pi q N}{n_{0}} \frac{\beta^{3}}{\omega_{0}^{8}}\left(\frac{6 a^{2}}{\omega_{0}^{2}}-3 b\right)
\label{eq1}
\end{aligned}
\end{eqnarray}
where the coefficients $a$, $b$, and $\beta$ can be expressed through the equations \cite{dolgaleva2015prediction}:

\begin{gather}
a = - \frac{m\omega^4_{0}a_{l}\alpha_{T}}{k_{B}},
b = \frac{6\pi q^{2}N\omega_{0}}{\hbar\left(n_{0,v}^2-1\right)}, \nonumber\\
\beta=\frac{\omega_{0}^{2}\left(n_{0,\nu}^{2}-1\right)}{4 \pi q N}
\label{eq2}
\end{gather}

Here $q$ denotes the effective charge of the chemical bond that represents the strength of the electrical coupling of the vibrational mode to the electric field of the radiation field, $N$ represents he concentration of oscillators as the number density of vibrational units, $n_{0}$ is the linear refractive index at the frequency of interest $\omega$, molecule diameter is $a_l$, the thermal expansion coefficient is $\alpha_T$, the vibrational part of linear refractive index is $n_{0,\nu}$. The complexity of the refractive indices is denoted by a bar above the symbols, which is omitted when referring to the real part of the $n_2$ coefficient. The conversion from esu units (denoted by tilde) to SI units (cm$^2/$W) holds according to the relation $n_2 = \frac{4 \pi}{3 n_0} \tilde{n}_2\times10^{-7}$\cite{boyd2008}.

It is important to note that in the original study \cite{dolgaleva2015prediction} the coefficient $b$ was assigned a positive sign, which was later corrected to negative, as confirmed by accurate numerical analysis \cite{guselnikov2023}. Nevertheless, the influence of the $b$-term on the calculated $n_2$ coefficient was found to be smaller than the experimental error, rendering it negligible for practical estimations \cite{guselnikov2023, nabilkova2023controlling}. Therefore, nonlinear coefficient $n_2$ can be can be simplified to: 
\begin{gather}
n_2\approx \frac{\pi q N}{n_{0}} \frac{6 a^{2}\beta^{3}}{\omega_{0}^{10}}.
\label{eq3}
\end{gather}
Previous studies on the nonlinear optical properties of liquids with complex structures, such as alcohols, 
demonstrated significantly enhanced nonlinearity indices compared to simpler molecules such as water \cite{tcypkin2021giant} across the THz, VIS, and IR spectral ranges. This enhancement arises from the presence of a richer vibrational spectrum with multiple polar bonds and differences in electronic and molecular properties. Unlike the water molecule, which has two equivalent O–H bonds, alcohols features a variety of bonds (O–H, C–O, C–H, C–C, etc.), each associated with distinct vibrational resonances \cite{sani2016spectral,wang2018temperature}. Therefore, Eq.~(\ref{eq1}) suitable for water \cite{tcypkin2019high}, should transform to into a summation over all over multiple vibrational modes ($\omega_{0, i}$) to account the molecular complexity of isopropanol. Such complicated approach for theoretical evaluation of nonlinearity demands specific knowledge of optical and physical characteristics attributed to each bond. However, there is other theoretical approach for estimation of the nonlinear refractive index coefficient of complex molecules based on known nonlinear properties of simpler molecules containing corresponding individual bonds.

To validate this approach, we considered isopropanol as a representative complex molecule. In this analysis, we use the established approach that claims vibrational nature of molecular nonlinearity dominance in the THz range. Specifically, we suppose that the fundamental stretching vibrational modes 1165, 2970 and 3000 cm$^{-1}$ corresponding to C-O and C-C combined peak, C-H, O-H are the primary contributors to isopropanol $n_2$ coefficient. The nonlinear refractive index quantifies the variation in a material's refractive index as a function of the intensity of incident light, adding $\Delta n$ to general refractive index $n_0$. Typically, the nonlinear refractive index values are significantly smaller than $n_0$. Therefore, it is reasonable to assume that nonlinear refractive index $n_2$ of isopropanol in THz range can be represented by summation of each bond contribution: $n_2^{O-H}$ + $n_2^{C-H}$ + $n_2^{C-O}$. 

We claim that the contribution of a specific oscillator, associated with a particular chemical bond in a complex molecule, can be derived from the $n_2$ of reference material with simpler structure composed predominantly of the same specific bond. The parameters $a$, $\omega_{0}$ and $\beta$ in Eq. \eqref{eq3}, as well as the original equation describing anharmonic oscillator behavior, are related only to the properties of the individual oscillator. In the case where inter-bond interaction is weak, the molecular polarization response to the electric field from same type of oscillator in different matters can be considered as equal. Based on these assumptions about equality of oscillator parameters, contribution $n_2^{O-H}$ to $n_2$ of isopropanol can be estimated through a nonlinear response $n_2^{H2O}$ of water \cite{tcypkin2021giant}, a molecule predominantly containing O–H bonds, according Eq. \eqref{eq3} using Eq. \eqref{eq3} adjusted by the concentration of O-H bond in the respective materials:

\begin{gather}
n_2^{O-H} = \frac{n_2^{H2O}N^{2prop}_{O-H}n^{H2O}_0}{N^{H2O}_{O-H}n^{2prop}_0}.
\label{eq4}
\end{gather}

Here, $n_2^{H2O}$ is measured nonlinear refractive index coefficient for water, $N^{2prop}_{O-H}$ is concentration of O-H bonds in isopropanol, $n^{H2O}_0$ linear refractive index of water, $N^{H2O}_{O-H}$ is concentration of O-H bonds in water and $n^{2prop}_0$ is linear refractive index of isopropanol equal to 1.48 \cite{huang2009improved}. The concentrations $N$ of bonds (the number density of vibrational units) are determined by material density and molecular composition: as the ratio of the specific gravity of material to the total mass of its molecule times the atomic mass unit (1.67$\times$10$^{-24}$). This value is multiplied by the number of bonds associated with this volume. The values used in these calculations are summarized in Table \ref{tbl1}, where experimentally obtained nonlinear refractive indices are denoted by (e) and theoretical estimations by (t). The resulting contribution of the O–H bond to the isopropanol nonlinear refractive index, normalized by concentration, is 1.3$\times$10$^{-10}$ cm$^2$/W. The similar approach (Eq. \eqref{eq4}) can be applied to evaluate contribution of $n_2^{C-H}$ to $n_2$ of isopropanol using the nonlinear response and corresponding concentration and refractive index parameters of $\alpha$-pinen, a molecule dominated by C–H bonds \cite{tcypkin2021giant,francis2020terahertz}. The calculated contribution of the C–H bond to the nonlinear refractive index of isopropanol, normalized by concentration, is 2.8$\times$10$^{-9}$ cm$^2$/W.

The estimation of $n_2^{C-O}$ contribution using the same methodology is more challenging due to the absence of experimental data for liquids composed purely of C–O bonds, for example carbon dioxide CO$_2$ in liquid form. A liquid-phase reference is necessary to avoid large discrepancies in concentration between gas and liquid phases. However, this contribution can be approximated using Eqs. (\ref{eq1}-\ref{eq2}) and available optical and physical data in THz and VIS ranges for liquid CO$_2$ under supercritical conditions \cite{wright1973density, marin2021ultraviolet}. It should be noted that such an estimation provides an upper bound, since under supercritical conditions, liquids exhibit gas-like diffusivity and liquid-like density which are maximize its light-matter interactions \cite{mareev2020optical}.

According to original research on critical CO$_2$ \cite{wright1973density} and subsequent studies \cite{marin2021ultraviolet} fundamental symmetric stretch mode $\nu_1$ of CO$_2$ under critical conditions -- temperature $T_c =$ 304.1 K, pressure $p_c =$ 73.8 bar and density $\rho_c =$ 480 kg m$^{-3}$ -- is observed at 1282 cm$^{-1}$ ($\omega_0 = $ 38 THz). This value aligns with the resonance peak in isopropanol associated with C-O (35 THz) and C-C (34 THz) stretching vibrations, which are combined and treated as a single vibrational mode with an effective fundamental frequency of 33 THz \cite{wang2018temperature}. The difference in resonance frequencies can be attributed to Raman shifts induced by bond interactions or coupling with other vibrations, as well as influence of critical conditions \cite{marin2021ultraviolet}. Under condensed-phase conditions, the thermal expansion coefficient $\alpha_{T}$ is reported as 0.01 1/K \cite{khonakdar2016mixed}. Unlike crystalline solids, where volume thermal expansion coefficient $\alpha_{T}$ shows how the matter expand under the influence of temperature in the well-defined lattice directions, liquids or gases lack long-range order, and their macroscopic $\alpha_{T}$ reflects the averaged expansion of all molecular bonds. For our estimation, we suppose that macroscopic value $\alpha_{T}$ of CO$_2$ is approximately equal to microscopic value $\alpha_{T}$ of C-O bond. The length of CO$_2$ molecule is approximately equal to twice the C-O bond length -- 2.32 \AA. Linear refractive index n$_0$ at 0.75 THz is 1.22 while electronic contribution $n_{0,el}$ is 1.108 that corresponds to value of linear refractive index in VIS range \cite{mounaix2003farzhang}. To determine vibrational contributions to the low-frequency refractive index $n_{0,\nu}^{(i)}$ for each resonance frequency, we employ the standard expression:  $n_{0}=\sqrt{1+\chi_{el}^{(1)} + \chi_{\nu}^{(1)}}$. Here $\chi_{el}^{(1)}$ and $\chi_{\nu}^{(1)}$ represent the electronic and vibrational susceptibilities. Other approximation parameters for $n_2$ evaluation are the following: the reduced mass of the vibration mode for A$_x$B$_y$ type of bond is calculated according to $\frac{m_A\times m_B}{m_A + m_B}$, where $m_A$ and $m_B$ denote the atomic masses of the respective elements. The specific gravity of CO$_2$ is equal to 0.48 according to its density. Utilizing these parameters, the predicted nonlinear refractive index $n_{2}$ for carbon dioxide CO$_2$ in liquid form in the low-frequency limit is calculated to be 5.8$\times$10$^{-9}$ cm$^2$/W. Consequently, $n_2^{C-O}$ contribution evaluated by using Eq. \eqref{eq4} is determined to be 1.8$\times$10$^{-9}$ cm$^2$/W. The contributions of all individual bond components, $n_{2}^{(i)}$, to the total nonlinear refractive index $n_2$ of isopropanol are summarized in Table \ref{tbl1}.

The Table \ref{tbl1} shows optical and physical parameters of reference materials containing predominantly a single bond type, which contribute to the overall nonlinear refractive index of isopropanol.

\begin{table*}
\caption{\label{tbl1}Isopropanol parameters for modeling.}
\begin{ruledtabular}
\begin{tabular}{ c c  c  c  c  c  c  c }
Contributing   & Reference & $\omega_0/2\pi$  & $N\times 10^{22}$ & $N\times 10^{22}$ & \multirow{2}{*}{$n_0$} & $n_{2}$ of ref. material & $n_{2}^{(i)} \times 10^{-10} $\\
bond &  material &THz &  ref. material & isopropanol  & &  $\times 10^{-10}$ cm$^2$/W &   cm$^2$/W\\
\hline
 O-H  & H$_2$O & 100 \cite{hale1973optical}         & 6.6  & 0.8 & 2.3 \cite{wilmink2011development} & 7 (e) \cite{tcypkin2019high} & 1.3 \\
 C-O  & CO$_2$ & 38 \cite{marin2021ultraviolet}     & 2.1  & 0.8 & 1.22 \cite{mounaix2003farzhang} & 58 (t) & 18\\
 C-H  & $\alpha$-pinen & 78 \cite{mayerhofer2023unveiling}  & 6.1  & 5.5 & 1.6 \cite{francis2020terahertz}  & 30 (e) \cite{tcypkin2021giant} & 28 \\
\end{tabular}
\end{ruledtabular}
\end{table*}

By model presented, theoretically predicted value of nonlinear refractive index $n_2$ of isopropanol is 4.7$\times$10$^{-9}$ cm$^2$/W.

\section{Experiment}
To confirm validity of the approach for $n_2$ evaluating, we use experimental framework. The nonlinear refractive index coefficient $n_2$ of isopropanol is experimentally determined using the Z-scan technique alongside with a pulsed THz radiation source. 

\begin{figure}
\centering\includegraphics[width=0.8\columnwidth]{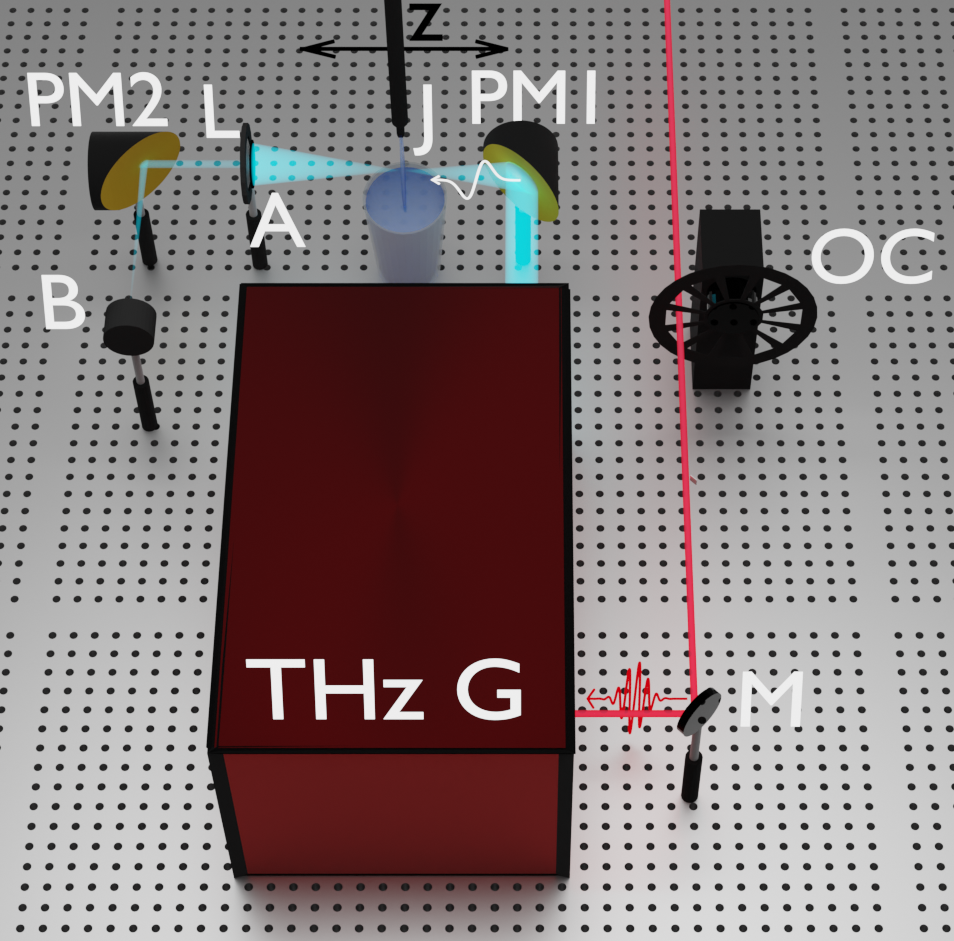}
\caption{
\label{FIG1}
Schematic experimental setup for the Z-scan measurements with depiction of liquid jet formation by nozzle \cite{watanabe1989new}. Here,  OC – optical chopper, M – mirror, THz G – terahertz radiation source, PM – parabolic mirror, J – jet, A – aperture, L – lens, B – bolometer.}
\end{figure}

Figure \ref{FIG1} shows the configuration utilized to ascertain the nonlinear refractive index $n_2$ of a planar liquid jet of isopropanol when subjected to terahertz (THz) pulse radiation.

The pulse, with a duration of 35 fs, has a central wavelength of 790 nm and a pulse energy of 1 mJ. It generated in a Ti:Sapphire femtosecond laser with a regenerative amplifier (Regulus Avesta project) and is guided into the terahertz generator (THz G) by mirrors (M). The pulse passes through an optical chopper (OC) modulator to synchronize it with detection module. The THz radiation source employed is the TERA-AX system (Avesta Project), which generates THz radiation through optical rectification of femtosecond laser pulses in a lithium niobate crystal (MgO:LiNbO$_3$) \cite{yang1971generation}. The resultant THz pulse is vertically polarized, possesses an energy of 400 nJ, 1 ps duration, and spans a spectral range of 0.1 to 2.5 THz with the maximum intensity at 0.75 THz ($\lambda$ = 0.4 mm) \cite{tcypkin2019high,tcypkin2021giant}. The THz beam's spatial size at the generator's output is 17.5 mm, with a caustic diameter of 1 mm (FWHM). The THz radiation is subsequently focused into a liquid jet (L) by parabolic mirror (PM1) with a 12.7 mm focal length. A short-focal parabolic mirror with a substantial numerical aperture is employed to achieve heightened intensity at the focal point. This configuration enables the attainment of peak radiation intensity within the THz beam's caustic, reaching $I_0$ = 0.5$\times$10$^8$ W/cm$^{2}$. To obtain the transmittance of the isopropanol jet, the radiation's average power is gauged using both open and closed apertures. A lens (L) is used for collimating the THz radiation. Aperture (A) is used for the Z-scan technique, which makes it possible to detect small beam distortions in the original beam. The linear transmission of the aperture $S$ is 2\%, which allows to maximize the sensitivity of the measurement method but reduces the signal-to-noise ratio. The second parabolic mirror (PM2) is used to focus the THz radiation into a bolometer (B) (Gentec-EO THZ5B-BL-DZ-D0), which is utilised to measure the power of the THz radiation.

The planar liquid jet is traversed along the caustic region from -3 mm to 3 mm (with geometrical focus at 0 mm using a motorized linear translator). The jet with a $L$ = 100 $\mu$m thickness is aligned perpendicularly to the incident radiation. The jet's displacement limitations are dictated by its width (8 mm) and the THz radiation's focusing geometry. The jet is formed by a nozzle that amalgamates a compressed-tube nozzle and a pair of razor blades, a design introduced by \cite{watanabe1989new}, creating a planar water surface (jet sheet) with plane-parallel flow shown in \cite{ismagilov2021liquid}. The THz pulse's optical path traverses the jet's central region, maintaining consistent thickness \cite{watanabe1989new}. The utilization of a pump facilitates liquid ejection under pressure, and the inclusion of a hydroaccumulator in the water supply system significantly mitigates pulsations associated with the water pump's operation \cite{ismagilov2021liquid}. 

\begin{figure}
\centering\includegraphics[width=1\columnwidth]{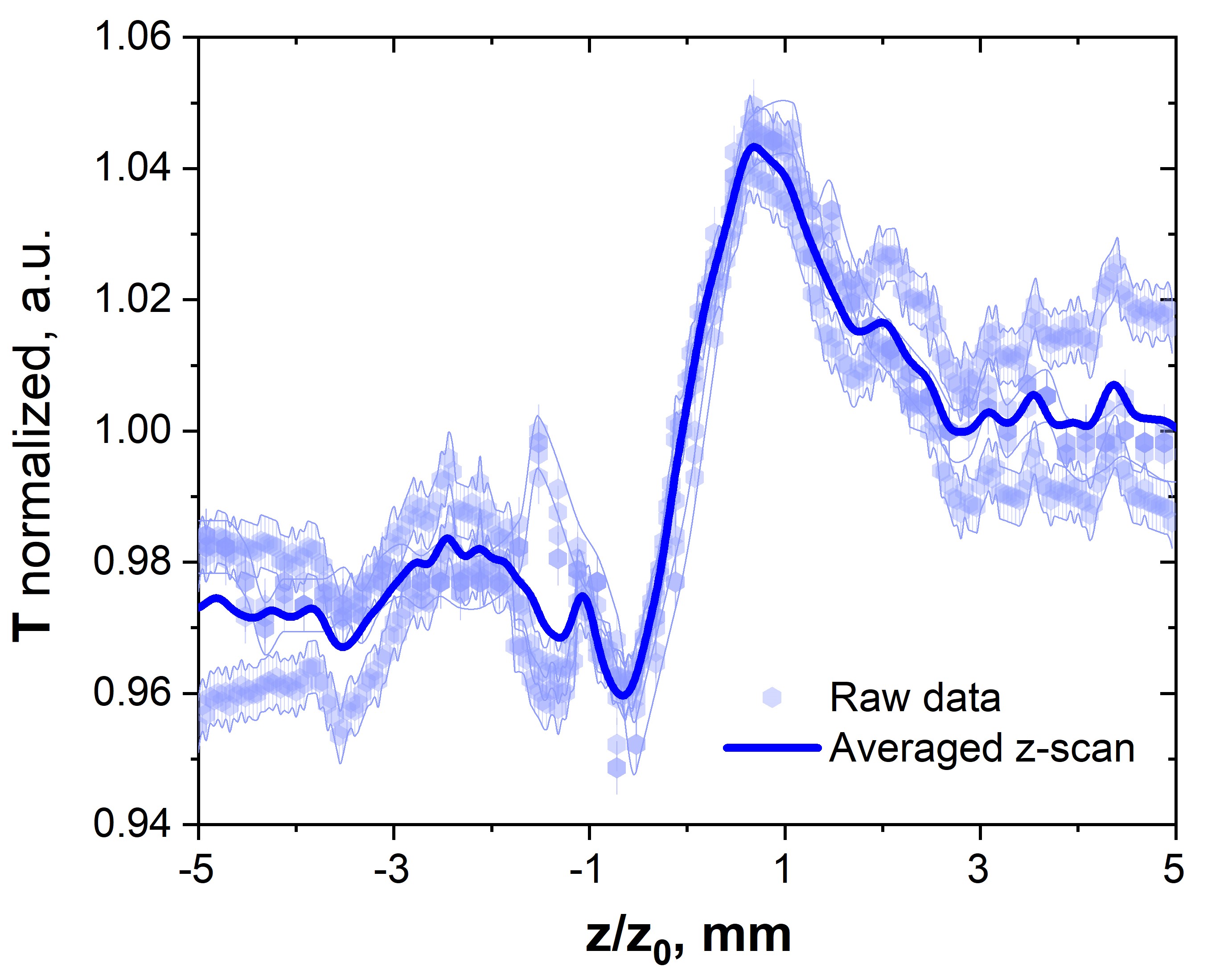}
\caption{
\label{FIG2}
Normalized Z-scan curve of isopropanol jet. Here, $z_0$ is Rayleigh length. Blue solid line stands for averaged data across various measurements.}
\end{figure}

Figure \ref{FIG2} shows the experimental results of the closed aperture Z-scan measurements normalized to the open aperture measurements. These experimental results demonstrate that the difference between the normalized peak and the valley transmittance $\Delta T$ = 0.1. Thus, using the empirical linear relationship between $\Delta T$ and nonlinear phase shift $\Delta \Phi_0$ \cite{sheik2002sensitive} we can find the nonlinear refractive index coefficient $n_2$ of isopropanol by the following equation:
\begin{equation}
n_2 = {\frac{\Delta T}{0.406I_{0}}\times \frac{\sqrt{2}\lambda}{2\pi L_\alpha (1-S)^{0.25}}}
\label{eq6}
\end{equation}
where $S$ is the linear transmission of the aperture equal to 0.02, $L_\alpha$= $\alpha^{-1}$ $\left[1-e^{-\alpha L}\right]$ is the effective interaction length, $L$ is the jet thickness, $\alpha$ is the absorption coefficient ($\alpha$ = 24 cm$^{-1}$ \cite{article24782}), $k = 2\pi/\lambda$ is the wave vector, $\lambda$ is the pulse central wavelength, and $I_{0}$ is the maximum input radiation intensity. Substituting the experimentally obtained value of $\Delta T$, we found $n_2$ =  2.5$\pm$0.5$\times$10$^{-9}$ cm$^2$/W. 

Theoretically obtained value of $n_2$ is slightly lager than experimental one. This can be attributed to upper bond evaluation of $n_2^{C-O}$ contribution. However, the agreement between the theoretical estimation and the experimentally obtained value suggests that the nonlinear response of complex molecules (such as isopropanol) can be accurately evaluated through a bond-specific contributions derived from simpler molecular systems (such as water, $\alpha$-pinen, CO$_2$).

\section{Conclusion}

This work provides a new theoretical approach for estimating the nonlinear refractive index $n_2$ of complex molecules with a high degree of accuracy. The isopropanol’s $n_2$ is used to demonstrate that the nonlinear optical response of complex molecules can be reliably approximated by summing the contributions of individual bonds, derived from structurally simpler molecular systems predominantly composed of a single bond type. By assuming that parameters, describing anharmonic oscillator, are specific to individual oscillators, the contributions of different molecular bonds to  $n_2$ can be treated independently under the assumption of additivity. The theoretical evaluation of isopropanol's $n_2$ obtained by summing the contributions $n_2^{O-H}$, $n_2^{C-H}$ and $n_2^{C-O}$ from $n_2$ of water, $\alpha$-pinen and CO$_2$, respectively, is found to be 4.7$\times$10$^{-9}$ cm$^2$/W. This theoretical prediction was corroborated through experimental validation, yielding a resultant value of 2.5$\pm 0.5 \times$10$^{-9}$ cm$^2$/W via pulsed THz Z-scan measurements at a central frequency of 0.75 THz. The consistency between the obtained nonlinear coefficient from experiment and the theoretical values predicted by the model supports the idea of quantitatively connection between microscopic bond vibrations and macroscopic nonlinear coefficients, providing a framework for understanding the nonlinear optical properties of molecular liquids.

The pronounced nonlinear response of isopropanol enables efficient manipulation of THz waves, opening up the possibility of using isopropanol-based systems in ultrafast optics: developing high-speed THz switches, modulators, and frequency converters. Furthermore, its large nonlinear refractive index allows the observation of nonlinear effects, such as self-focusing, Kerr effect, and THz harmonic generation, at much lower laser intensities compared to materials commonly used in the visible and infrared ranges. The findings of this study may pave the way for designing advanced photonic devices and systems. The ability to exploit the ultrafast optical response of isopropanol and similar liquids in THz technology will drive innovations in THz-based communication, imaging, and ultrafast signal processing.

\bibliography{apssamp}

\end{document}